\documentclass[
twocolumn,
]{ceurart}

\usepackage{todonotes}
\usepackage[inline,shortlabels]{enumitem}
\usepackage{graphicx}
\usepackage{subfig}
\usepackage{hyperref}
\usepackage{url}
\usepackage[most]{tcolorbox}
\usepackage{fontawesome5}

\usepackage[nameinlink]{cleveref}
\crefname{figure}{Fig.}{Figs.}
\Crefname{figure}{Figure}{Figures}
\crefname{equation}{Eq.}{Eqs.}
\Crefname{equation}{Equation}{Equations}
\crefname{section}{§}{§§}
\Crefname{section}{Section}{Sections}
\crefname{table}{Table}{Tables}
\crefname{appendix}{Appendix}{Appendices}

\usepackage{color}

\usepackage{color}

\usepackage{listings}

\newcommand{\mcomment}[1]{}

\usepackage{colortbl}

\begin{document}

\copyrightyear{2024}
\copyrightclause{Copyright for this paper by its authors.
  Use permitted under Creative Commons License Attribution 4.0
  International (CC BY 4.0).}

\conference{RecSys in HR'24: The 4th Workshop on Recommender Systems for Human Resources, in conjunction with the 18th ACM Conference on Recommender Systems, October 14--18, 2024, Bari, Italy.}

\title{On the Biased Assessment of Expert Finding Systems}

 \author[1,2]{Jens-Joris~Decorte}[%
 email=jensjoris@techwolf.ai,
 url=https://www.techwolf.ai,
 ]
 \cormark[1]
 \address[1]{Ghent University -- imec,
   9052 Gent, Belgium}
 \address[2]{TechWolf,
   9000 Gent, Belgium}

 \author[2]{Jeroen~Van~Hautte}[%
 email=jeroen@techwolf.ai,
 url=https://www.techwolf.ai,
 ]

 \author[1]{Chris~Develder}[%
 email=chris.develder@ugent.be,
 ]

 \author[1]{Thomas~Demeester}[%
 email=thomas.demeester@ugent.be,
 ]

\cortext[1]{Corresponding author.}

\begin{abstract}
In large organisations, identifying experts on a given topic is crucial in leveraging the internal knowledge spread across teams and departments. 
So-called enterprise expert retrieval systems automatically discover and structure employees' expertise based on the vast amount of heterogeneous data available about them and the work they perform. 
Evaluating these systems requires comprehensive ground truth expert annotations, which are hard to obtain.
Therefore, the annotation process typically relies on automated recommendations of knowledge areas to validate.
This case study provides an analysis of how these recommendations can impact the evaluation of expert finding systems. 
We demonstrate on a popular benchmark that system-validated annotations lead to overestimated performance of traditional term-based retrieval models and even invalidate comparisons with more recent neural methods. 
We also augment knowledge areas with synonyms to uncover a strong bias towards literal mentions of their constituent words. 
Finally, we propose constraints to the annotation process to prevent these biased evaluations, and show that this still allows annotation suggestions of high utility. 
These findings should inform benchmark creation or selection for expert finding, to guarantee meaningful comparison of methods. 
\end{abstract}


\maketitle

\section{Introduction}
As organisations grow in size, effectively leveraging internal expertise becomes harder, as it is harder to locate experts on specific matters. 
This need for efficiently locating expertise in large organisations has been recognized for over two decades, with early systems such as \textit{P@NOPTIC Expert} that automatically identify experts based on information in an organisation's intranet~\cite{craswell2001p}.
This task, known as expert finding, is a specialized form of information retrieval (IR) where the focus is on identifying individuals with relevant expertise rather than on retrieving documents.
Another related task, expert profiling, focuses on retrieving all areas of expertise for a given individual~\cite{balog2007expertprofiling}.
The evaluation of these tasks, grouped under \emph{expertise retrieval}~\cite{balog2012expertise}, requires developing a comprehensive gold standard of expertise annotations. 
A setup where a list of experts is annotated for a given topic proves difficult, resulting in either just one or two experts linked per topic~\cite{craswell2005overview,ian2008overview}, or in the case of more extensive expert lists, limited to a mere seven topics overall~\cite{deng2008formal}.
As a result, annotation efforts have shifted towards asking the experts themselves to list their areas of expertise~\cite{balog2007broad}. 
As expert profiling and expert finding are strongly related, there is a trend of using expert profiling benchmarks for the task of expert finding as well~\cite{balog2007broad,berendsen2013assessment,mangaravite2016lexr}.
If a complete and accurate gold standard of expertise profiles is available, it can be inverted to identify relevant experts for a specific topic. 
However, achieving such comprehensive and accurate profiles is often unrealistic. 
The sheer number of topics typically precludes exhaustive consideration during annotation.
Secondly, self-selected topics rely on the expert’s recollection and understanding of the system’s taxonomy, often resulting in sparse profiles subject to cognitive biases, such as recency bias.
To address this, the annotation process often includes an automated system recommending additional, likely topics for each expert~\cite{berendsen2013assessment,mangaravite2016lexr}.
These system-validated topics yield more comprehensive and varied expert profile annotations.

Note that these personalized recommendations can greatly influence which topics each expert might consider during annotation. 
We argue that this can, under certain circumstances, preclude meaningful comparisons of annotations across experts, and therefore the use of these benchmarks for expert finding systems. 
Specifically, this work addresses these research questions:
\begin{itemize}[leftmargin=*]
    \item \textbf{RQ1:} Do system-validated and self-selected annotations exhibit significantly different characteristics that could bias the evaluation of expert finding systems?
    \item \textbf{RQ2:} How do system-validated annotations impact perceived performance of term-based versus neural retrieval systems in expert finding tasks?
    \item \textbf{RQ3:} Can we establish constraints for a new annotation setup that ensures the evaluation of expert finding systems remains representative and unbiased?
\end{itemize}

We address these questions based on an analysis of the popular TU Expert Collection~\cite{berendsen2013assessment}, which makes available multiple sets of ground truth that nicely facilitate our analysis. 
The nature of self-selected and system-validated expertise profiles is compared, specifically examining the properties of system-validated annotations and their impact on the validity of the expert finding task in section~\ref{sec:analysis}. 
We implement both traditional term-based retrieval systems and more recent neural IR methods in section~\ref{sec:assessment}. 
Additionally, the section covers a procedure to augment all test queries in the TU Expert Collection with synonyms, allowing to further analyze the effect of any term-based biases in the annotations. 
We also propose constraints for system-validated annotations and demonstrate the potential of a new annotation suggestion system in this section. 
Finally, section~\ref{sec:discussion} discusses all results and provides answers to the research questions.

\section{Related Work}

\paragraph{Expert finding systems}
The development of expert finding systems has a rich history, starting with the introduction of \textit{P@NOPTIC Expert}, an early systems designed to automatically identify experts based on textual documents available within an organisation's intranet~\cite{craswell2001p}. 
To this day, expert finding remains an important and challenging topic for large organisations across different niches like the medical domain~\cite{sahar2024artificial}. 
Most expert finding methods are formalized as one of two prominent models: the candidate model and the document model, referred to as the query-independent and query-dependent models, or Model 1 and Model 2, respectively~\cite{balog2006formal}. 
The candidate model regards each candidate expert as the set of their linked documents, to directly retrieve relevant experts given a query. 
The document model operates on two steps, first determining the relevance of individual documents towards a query, and afterwards aggregating this document ranking into a candidate ranking. 
Of these, the document model (Model 2) has generally been shown to be more effective~\cite{balog2006formal}. 
Subsequent research has largely built upon these models, with some studies exploring expert finding as a voting problem, utilizing data fusion techniques in a metasearch framework~\cite{macdonald2006voting}.
Other works have an extended input data scope to the expert finding system, such as by incorporating prior topic distributions~\cite{fang2007prob} or by leveraging document structure to enhance retrieval performance~\cite{zhu2006theopen}. 
Our work relies on the presence of a textual corpus linked to employees without further properties like document structure.

\paragraph{Expert finding benchmarks}
The first large-scale benchmarks for expert finding were developed as part of the TREC Enterprise track, which ran from 2005 to 2008. 
It introduced the W3C expert corpus in 2005, alongside an \textit{expert search} task consisting of 50 knowledge areas, annotated with experts from a total of 1,092 candidates~\cite{craswell2005overview}. 
In 2007, the CERC dataset was introduced~\cite{ian2008overview}. 
Similar to the W3C dataset, it included 50 topics, which were developed by nine science communicators at CSIRO. 
These communicators were then tasked with identifying one or two CSIRO staff members as experts on each topic, contributing to a robust dataset for expert finding research.
Another notable dataset is derived from DBLP biographical data, augmented with abstracts from Google Scholar~\cite{deng2008formal}. 
This dataset contains 953,774 papers in total and 574,369 valid authors, with 2,498 topics sourced from a research events website. 
However, only seven topics have been annotated with expert lists, each containing between 20 and 45 experts~\cite{deng2008formal}. 
The dataset development process, both for TREC and the DBLP-based dataset, highlights the difficulty of identifying experts on a certain topic within large organisations, as annotators often lack detailed knowledge of the topics themselves.

\paragraph{Expert profiling benchmarks}
A more distributed approach to gathering annotations involves asking employees to fill in their own expertise profiles, as seen in the UvT dataset~\cite{balog2007broad}. 
This new annotation scheme became more prominent shortly after the introduction of the task of expert \textit{profiling} (rather than \textit{finding}), where the goal is to retrieve all areas of expertise for a given individual~\cite{balog2007expertprofiling}. 
The UvT dataset was the first large-scale benchmark for expertise profiling relying on self-reported expertise. 
However, this approach often results in sparse profiles due to the difficulty of recalling all areas of expertise.
To address this sparsity, semi-automated annotation procedures have been proposed. 
Berendsen et al.\cite{berendsen2013assessment} extended the self-selected profiles from the UvT dataset by presenting up to 100 high-probability topics for further annotation, re-releasing the new annotation sets under the name of the TU Expert Collection. 
Similarly, Mangaravite et al.~\cite{mangaravite2016lexr} employed a content-based tag recommendation system to suggest annotations for approval. 
These enriched annotations, while originally intended for expert profiling, have been increasingly used to evaluate expert finding tasks as well~\cite{gysel2016unsupervised,CIFARIELLO20191}. \\

\noindent The use of personalized annotation suggestions raises concerns about the validity of these annotations for evaluating expert finding systems, which this study aims to address.
Specifically, because these suggestions are different for each employee, the impact of this mechanism may introduce properties in the annotations that forego their comparability across employees.
Our work is closely related to that of Berendsen et al.~\cite{berendsen2013assessment}, who conducted an extensive study on the impact of different expert profile annotation schemes on the evaluation of expert profiling tasks. 
However, our focus diverges in that we specifically investigate the impact of these benchmarks on expert finding, uncovering significant challenges in their usability under system-validated setups.

\section{Analysis of Annotation Schemes}\label{sec:analysis}

We analyze the differences of self-selected versus system-validated expertise annotations, and how they may influence the perceived performance of expert finding systems when used as a benchmark. 
Section~\ref{sec:tu} introduces the TU Expert Collection, which is the dataset used in this analysis, as introduced in \cite{berendsen2013assessment}.
We perform initial analysis of the annotation suggestions in section~\ref{sec:distr}, showing their utility for expanding profile annotations, but also their high false negative rate. 
Finally, section~\ref{sec:termbias} analyzes the mechanism behind these false negatives, exposing a large positive bias towards literal mentions of the knowledge topics' constituent words in the corpus. 

\subsection{TU Expert Collection}\label{sec:tu}

The TU expert collection is an expertise retrieval benchmark focused on a knowledge-intensive organisation, namely the Tilburg University~\cite{berendsen2013assessment}.
It is an updated version of the earlier UvT dataset~\cite{balog2007broad}.
The dataset contains a variety of documents, being academic publications, supervised student dissertations, course descriptions, and research summaries. 
These documents are primarily in Dutch and English, and are explicitly linked to experts in the university's Webwijs system, indexes over 2,000 unique knowledge areas and 761 employees.
The TU dataset provides several ground truth (GT) sets of graded expert profile annotations, labeled GT1 through GT5.
These annotations are the result of experts indicating their expertise areas on a scale of 1 (lowest) to 5 (highest). 
Note that in this work, we consider all annotations as binary and ignore the attached grades, due to inconsistencies in how different annotators may interpret and apply these grades, compromising the comparability of grades across experts~\cite{berendsen2013assessment}. 
We leave the analysis of graded relevance annotations for future work. 
GT1 contains the self-selected knowledge areas of 761 TU employees. 
To further expand these annotations, a system was developed to recommend up to 100 highly likely knowledge areas to each expert, which they could easily validate or discard through a user interface. 
The extended expertise profiles of all 239 participating employees is bundled as GT5, accompanied by GT4 which simply binarizes the annotations by dropping the graded relevance scores. 
For ease of use, GT2 is provided as the subset of GT1 for those 239 employees.
Finally, GT3 is an alteration of self-selected profiles in GT2, reduced to only those topics that were also present in the personalized suggested annotations. 

\subsection{Distribution of system-validated annotations}\label{sec:distr}

As reported in \cite{berendsen2013assessment}, the average self-selected profile in GT1 contains 6.4 knowledge areas. 
The average size of the extended profiles in GT5 expands to 8.6 areas. 
Notably, we find that the percentage of employees with three or fewer expertise areas decreases from 19.7\% in GT1 to 10.5\% in GT5.
Additionally, system-generated profiles capture 81\% of the final knowledge areas compared to 65\% for self-selected profiles, and unique topics in the annotations grows from 937 in GT1 to 1,266 in GT5. 
This shows the sparsity of self-selected expertise profiles, and how the situation improves through personalized system recommendations for annotation. 
However, because these recommendations are personal to each expert, the underlying recommendation method may compromise the comparability of the annotated knowledge areas across experts. 
To explore this, we focused on niche topics, present in more than one but no more than three self-selected profiles in GT2, identifying 290 such topics. 
Examples are \texttt{sub-saharan africa}, \texttt{policy evaluation}, \texttt{cognitive linguistics}, \texttt{nonprofit organisations} and \texttt{extreme value theory}. 
By contrasting GT2 with GT3, we know whether a self-selected topic was also part of the system annotation suggestions, allowing us to estimate the recall of this system. 
We find that only 125 out of the 290 niche topics -- around 43\% -- were recommended for annotation to all experts who had self-selected it.
It is this low recall that can compromise the comparability of annotations across experts: if it is caused by a certain weakness of the annotation recommendation system, expert finding systems with a similar weakness will produce the same recall patterns and therefore may appear stronger than they are.  

\subsection{Term-bias in system-validated annotations}\label{sec:termbias}

To construct GT5, up to 100 knowledge areas were suggested for further annotation to each expert, produced by an ensemble of eight expert profiling systems~\cite{berendsen2013assessment}. 
These systems vary in retrieval models (Model 1 or Model 2), the query language (English or Dutch), and whether they consider relationships between topics in the Webwijs system. 
All these systems have in common that they model the probability of a topic for a document or expert based on the literal textual occurrences of their constituent words. 
This approach is prone to false negatives due to its inability to account for synonyms and other semantic nuances, leading to a low recall and a strong bias towards literal mentions of the topic's constituent words. 

\begin{figure}[ht]
\centering
\includegraphics[width=\columnwidth]{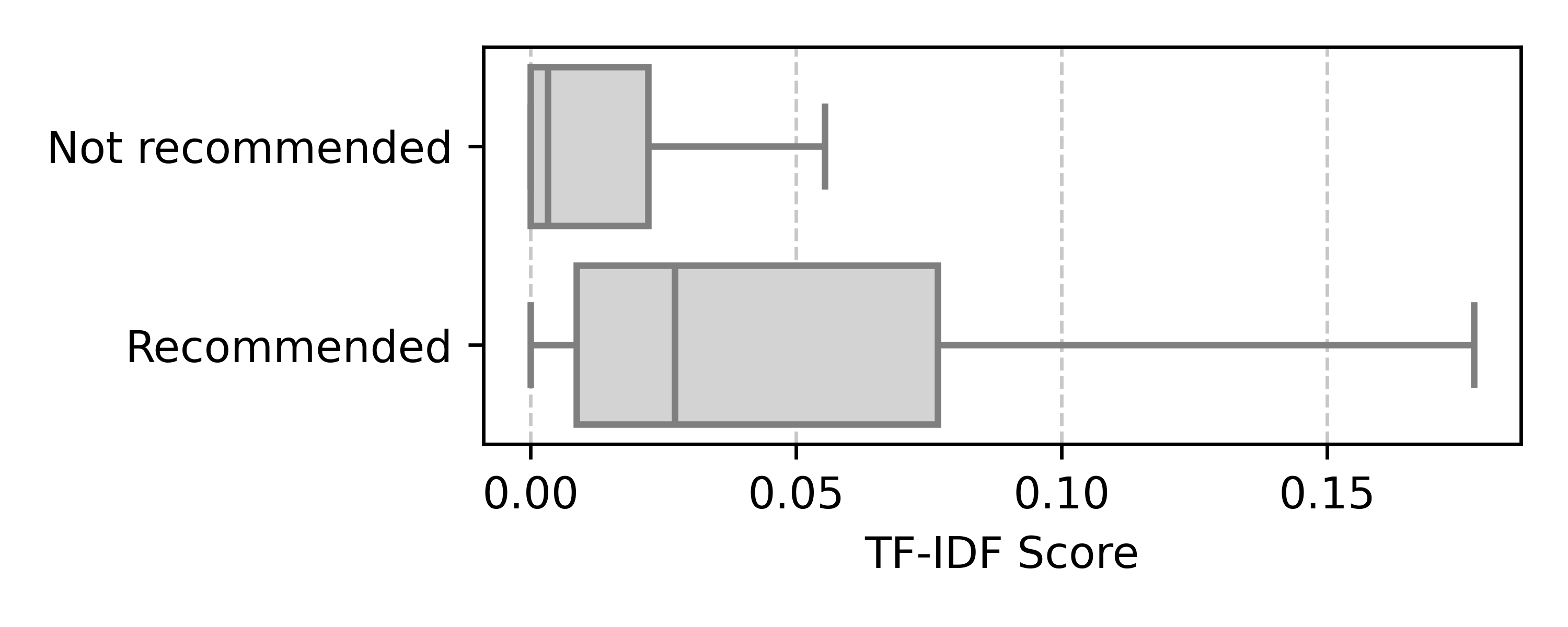}
\caption{Distribution of tf-idf scores for self-selected topics that were part of the annotation suggestions versus those that were not. Significantly higher scores are observed for those that were part of the annotation suggestions.}
\label{fig:recornot}
\end{figure}

We aim to quantify the presence of this bias towards literal mentions in system-validated topics or their constituent words. 
To this end, we construct a corpus with one long document per expert, being the concatenation of all original documents linked to the expert. 
We then calculate tf-idf scores of queries with respect to these concatenated expert documents to express the degree to which their constituent words are being literally mentioned. 
Whenever both an English and Dutch name are available for a topic, we consider the largest of the tf-idf scores of both versions. 
By contrasting GT2 and GT3, we are able to determine for each topic in the self-selected profiles whether it was also included in the suggested annotation list for the corresponding expert. 
Figure~\ref{fig:recornot} shows the distribution of tf-idf scores for both groups of self-selected topics through a boxplot, with commonly used whiskers drawn to the farthest datapoint within 1.5 times the interquartile range from the nearest hinge. 
It shows a large shift in scores for the subset of self-selected topics that are also included in the annotation recommendations versus those that are not included.

\begin{figure}[ht]
\centering
\includegraphics[width=\columnwidth]{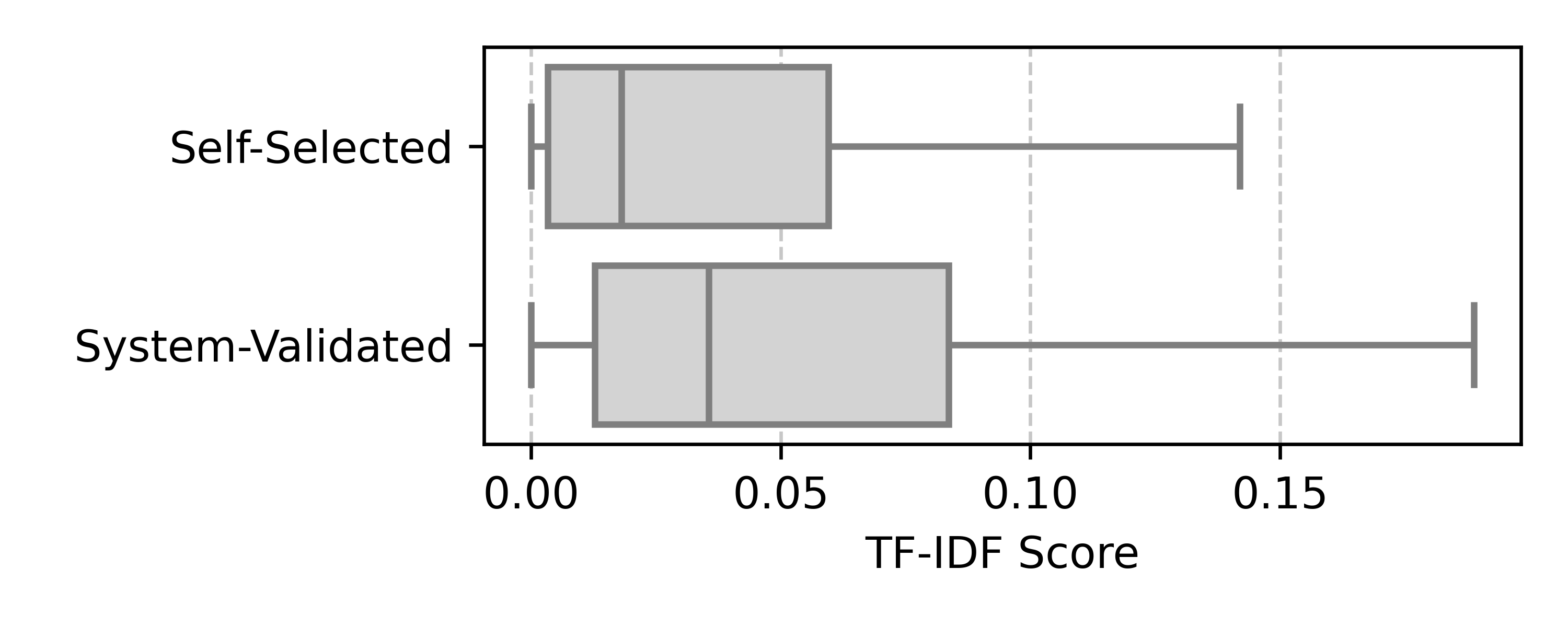}
\caption{Distribution of tf-idf scores for self-selected topics versus additional topics added through system validation.}
\label{fig:selfvsval}
\end{figure}

Given this observation, we expect this bias to be extended into the additional topics that are added to the profiles through system-validation. 
Figure~\ref{fig:selfvsval} shows the distribution of tf-idf scores of the topics in the original self-selected expertise profiles, compared to the topics added after system-validation. 
As expected, a clear positive bias in tf-idf scores is observed for the system-validated topics. 
These findings validate our concerns about system-driven annotation biases.

In conclusion, our analysis reveals a strong term frequency bias in the annotation recommendations. 
As a result, there are likely false negatives in the system-validated annotations that systematically favor systems with similar term frequency-based biases. 
This suggests that evaluations of expert finding systems using these annotations may overestimate the performance of systems that share these biases.

\section{Assessment of Expert Finding}
\label{sec:assessment}

\subsection{Expert Finding Models}

We implement two different expert finding systems and evaluate them against different ground truths.
Both systems operate under the query-dependent Model 2, first determining the relevance of each document in the corpus to the query and then aggregating document ranks into a final ranking of experts. 
Two different information retrieval models are implemented to rank the documents against a query. 
We implement a popular term-based retrieval model, as well as a more modern neural IR technique, aiming to maximally expose potential term-based biases in the annotations.

\paragraph{Term-based retrieval:}
we use BM25~\cite{robertson1994some} to rank all documents against the query topic because it is a commonly used term-based retrieval method that accounts for document lengths. 
In order to support the multi-linguality of the dataset, we simply concatenate the English and Dutch translations of the topic when both are present.

\paragraph{Late-interaction neural retrieval:} 
We opt for the ColBERT retrieval technique~\cite{omar2020colbert}, due to its unique combination of efficiency and strong IR capabilities. 
To facilitate the multilingual nature of the corpus, we use the multilingual ColBERT-XM model~\cite{louis2024modular}. 
The corpus is split into chunks of up to 256 tokens, with an overlap of 64 tokens. 
We make use of the \textit{RAGatouille}\footnote{\url{https://github.com/AnswerDotAI/RAGatouille}} implementation to chunk and index the whole corpus, which took close to eight hours on one NVIDIA P100 GPU. 
As with the BM25 model, we simply concatenate the English and Dutch translations of the topic when present. 
The documents are ranked according to the average retrieval score of all their constituent chunks. 

\bigbreak

\noindent Following \cite{CIFARIELLO20191}, for both IR methods, we use the same \textit{rr} function to aggregate the document ranking into an expert ranking, as defined by:

\[
\text{rr}(c) = \sum_{j=1}^{|D_{c,q}|} \frac{1}{\text{rank}(d_j)}
\]

In this equation, $D_{c,q}$ is the subset of documents linked to candidate $c$ that are retrieved for query $q$.
The rank of the retrieved documents is indicated by $\text{rank}(d_j)$, starting at $1$ for the highest ranked document.

\subsection{Query Augmentation}

With the goal of further studying the impact of term-based biases in the annotations, we extend the Webwijs knowledge areas with synonyms. 
Qualitative expert finding system should surface relevant experts on a topic, even if provided by a synonym of the query instead of the original query.
Because of this, query synonyms provide an opportunity to study term-based annotation biases. 
We manually annotate $109$ randomly selected knowledge areas with both English and Dutch synonyms. 
To facilitate the annotation process, the topic up for annotation is contextualized by providing the annotator with all its relations to other topics in the Webwijs inventory. 
For further contextualization, we gather the corresponding wikidata and wikipedia page on the topic if they exist. 
Whenever available, good synonyms are selected from the \textit{``Also known as''} table in wikidata. 
If not, we scan the first paragraph of the wikipedia page for synonyms. 
Finally, we manually provide a synonym if none are available in the wiki pages.
An example result is shown here: 

\begin{tcolorbox}[fontupper=\footnotesize, colback=lightgray!20, boxrule=0.5pt, arc=4pt, boxsep=0pt, left=4pt, right=4pt, top=2pt, bottom=2pt, colframe=black, sharp corners]
\texttt{\underline{Topic EN:}
auction theory\\
\underline{Topic NL:}
veilingstheorie\\
\\
\underline{Webwijs links:}\\
\textit{Makes use of:} auctions / veilingen\\
\\
\textbf{Annotation:}\\
\underline{wikidata:} \url{https://www.wikidata.org/wiki/Q771334}\\
\underline{wikipedia:} \url{https://en.wikipedia.org/wiki/Auction_theory}\\
\\
\underline{Synonym EN:}
Bidding Theory\\
\underline{Synonym NL:}
Biedingstheorie
}
\end{tcolorbox}

\noindent Based on the initial set of synonym annotations, we automatically generate synonyms for all remaining queries using OpenAI's GPT4o model, selected for its high accuracy on common benchmarks.
We randomly select 35 annotated queries for training and the remaining 74 for validation, and automatically optimize a chain-of-thought prompt~\cite{wei2022chain} using DSPy's \texttt{BootstrapFewShot} prompt optimization technique~\cite{khattab2023dspy}.
We release the full set of synonym annotations.\footnote{\url{https://huggingface.co/datasets/jensjorisdecorte/TU-Expert-Collection-Topic-Synonyms}}
Limited manual quality checks were performed on these synonyms.
We note that this process can be more qualitatively performed in future studies, however the main reasoning of using topic synonyms to indicate a bias towards literal mentions of their constituent words still holds, irrespective of suboptimal quality of the synonyms.

\subsection{Alternative Annotation Suggestions}

We analyze an alternative approach to suggest additional knowledge areas for annotation. 
Specifically, we introduce the constraint that the suggestion mechanism has no access to the document corpus, and may only recommend additional knowledge areas based on the available self-selected topics. 
This setup prevents any systematic bias stemming from the annotation procedure with respect to the textual corpus. 
The downside of this setup is that the rich information in the corpus cannot be utilised, and it instead relies on at least a small number of manually annotated expertise areas per expert. 
We argue that experts should be sufficiently engaged in the annotation process such that this is a reasonable requirement. 
Additionally, there is an opportunity for these annotation suggestions to dynamically adapt throughout the validation procedure, although we leave this out of the current scope. 

We develop two item-to-item recommendation systems (where the Webwijs topics serve as items).
The first system learns from the full set of self-selected profiles in GT1, and recommends topics that have high pointwise mutual information (PMI) with the given topic.
PMI is a simple but effective measure of association between items. 
Topics that occur three times or less are excluded from this system to ensure a minimum level of robustness.
The second system operates on all Webwijs topics, and recommends topics that are semantically similar to a given topic, as measured by the cosine similarity of semantic neural embeddings of their respective names. 
We use a generic multilingual sentence-transformer model (\textit{paraphrase-multilingual-mpnet-base-v2})\footnote{\url{https://huggingface.co/sentence-transformers/paraphrase-multilingual-mpnet-base-v2}} that was pretrained on over 1B English sentence pairs~\cite{reimers-gurevych-2019-sentence} and then adapted to over 50 languages through a knowledge distillation process~\cite{reimers-gurevych-2020-making}.
We embed English and Dutch names separately and always consider the highest similarity between either versions.
For a given self-selected profile, we compile a list of up to 100 annotation recommendations by pooling the top recommendation of both systems, looping over the self-selected expertise topics in a round-robin fashion.

\section{Results and Discussion}\label{sec:discussion}

\begin {table*}[ht]
\begin{tabular}{lcccccccccccc}\toprule
& \multicolumn{4}{c}{\textbf{GT2}} & \multicolumn{4}{c}{\textbf{GT5}} & \multicolumn{4}{c}{\textbf{GT5} with synonym queries}
\\\cmidrule(lr){2-5}\cmidrule(lr){6-9}\cmidrule(lr){10-13}
                & P@5 & MAP & nDCG & MRR & P@5 & MAP & nDCG & MRR & P@5 & MAP & nDCG & MRR \\\midrule
\emph{BM25}     & 15.47 & 37.81 & 52.39 & 46.06 & \textbf{21.94} & \textbf{56.56} & \textbf{67.53} & \textbf{64.02} & 12.16 & 31.68 & 46.41 & 38.12 \\
\emph{ColBERT}  & \textbf{16.29} & \textbf{39.78} & \textbf{54.07} & \textbf{48.43} & 18.44 & 46.46 & 59.39 & 54.41 & \textbf{13.98} & \textbf{34.21} & \textbf{48.68} & \textbf{40.42}\\
\bottomrule
\end{tabular}
\caption{Performance metrics for both the term-based (BM25) and the late-interaction neural retrieval based (ColBERT) methods under different annotation schemes. GT2, containing just self-selected expertise topics, shows stronger performance for the ColBERT method, while the addition of system-validated topics in GT5 leads to a strong edge for the BM25 system. Using the same GT5 annotations, but swapping test queries for their synonyms, yields a system ranking that is again in accordance with GT2.}
\label{table:results}
\end{table*}

\paragraph{Impact of the annotation on expert finding evaluations}

We evaluate system performance using precision at 5 (P@5), mean average precision (MAP), normalized discounted cumulative gain (nDCG) and mean reciprocal rank (MRR). 
These metrics provide a comprehensive view of the retrieval system’s ability to rank relevant experts at the top of the list.
All evaluations are conducted using the official TREC evaluation software\footnote{\url{https://trec.nist.gov/trec_eval/}}, ensuring standardized and comparable results.
We report the performance of the BM25 and the ColBERT system on both the self-selected profiles (GT2) as well as their larger counterparts extended with system-validated topics (GT5) in table~\ref{table:results}.
We find that the ColBERT-based expert finding system outperforms its BM25-based counterpart on all metrics when using the self-selected profiles (GT2) as ground truth. 
The performance of both systems increases considerably when assessing them against the system-validated profile (GT5).
However, the increase in performance is much more drastic for the BM25-based system, leading it to strongly exceed ColBERT.
We hypothesize that it is the term-frequency bias in the annotation procedure of GT5 that leads to a strongly overestimated performance measurement for the term-based BM25 approach.
When swapping the test queries in GT5 for their synonyms, a significant drop in performance is observed, especially for the BM25 system which drops over 20 \%-points in MAP, nDCG and MRR.
We also observe that system ranking in this scenario corresponds to that under the GT2 evaluation. 

\paragraph{Statistics of the alternative annotation suggestions}

We perform the same analysis as in section~\ref{sec:analysis} with respect to the recall of 290 specific topics in the annotation recommendations.
This requires the specific topic for which recall is measured to first be removed from the self-selected profile.
The results show that 235 out of the 290 specific topics (around 81\%) is being recommended to every expert that had self-selected the topic.
Compared to the 43\% of the content-based annotation recommendations in the original study, this is a considerable increase in recall, which should further improve the comparability of annotations across experts. 
For completeness, we also report how well these new recommendations cover the topics that where added in GT5 compared to GT2.
We find that a total of 1,059 topic additions are made, and 505 (around 48\%) of these topics are also present in our proposed annotation recommendation method.
Because this proposed method has no access to the documents in the organisation, we do not expect strong overlap, and we consider 48\% to be relatively high.
Apart from this overlap measurement, it is difficult to assess the true precision of these recommendations because we did not have access to the candidate experts to facilitate manual validation. 
However, examples of the topic recommendations is provided in appendix~\ref{app:rec-examples}.

\bigbreak
\noindent In conclusion, our analysis of the TU Expert Collection allows us to answer \textit{yes} to all three research questions. 
Section~\ref{sec:analysis} provides an answer to \textbf{RQ1}, showing that the underlying mechanism used for the system-validated topics is subject to a high false negative rate, and that it introduces a significant bias towards literal mentions of knowledge areas' constituent words. 
With respect to \textbf{RQ2}, as shown above, the perceived performance of expert finding systems is indeed strongly impacted by these system-validated topics, and they even lead to significant differences in the ranking of these systems. 
Finally, we have proposed an annotation suggestion procedure that is independent of the document corpus, and we have developed such a system accordingly. 
It exhibits strong utility for the annotation process while significantly reducing the false negative rates observed in the original benchmark, leading us to answer \textbf{RQ3} positively as well.
Our analysis should help future work on expert finding -- or evaluation thereof -- make more informed decisions with respect to the selection or creation of these benchmarks.

\begin{acknowledgments}
We thank the anonymous reviewers for their valuable feedback. This project was funded by the Flemish Government, through Flanders Innovation \& Entrepreneurship (VLAIO, project HBC.2020.2893).
\end{acknowledgments}

\newpage

\bibliography{paper.bib}

\begin{thebibliography}{23}
\expandafter\ifx\csname natexlab\endcsname\relax\def\natexlab#1{#1}\fi
\providecommand{\url}[1]{\texttt{#1}}
\providecommand{\href}[2]{#2}
\providecommand{\path}[1]{#1}
\providecommand{\DOIprefix}{doi:}
\providecommand{\ArXivprefix}{arXiv:}
\providecommand{\URLprefix}{URL: }
\providecommand{\Pubmedprefix}{pmid:}
\providecommand{\doi}[1]{\href{http://dx.doi.org/#1}{\path{#1}}}
\providecommand{\Pubmed}[1]{\href{pmid:#1}{\path{#1}}}
\providecommand{\bibinfo}[2]{#2}
\ifx\xfnm\relax \def\xfnm[#1]{\unskip,\space#1}\fi
\bibitem[{Craswell et~al.(2001)Craswell, Hawking, Vercoustre, and Wilkins}]{craswell2001p}
\bibinfo{author}{N.~Craswell}, \bibinfo{author}{D.~Hawking}, \bibinfo{author}{A.-M. Vercoustre}, \bibinfo{author}{P.~Wilkins},
\newblock \bibinfo{title}{P@ noptic expert: Searching for experts not just for documents},
\newblock in: \bibinfo{booktitle}{Ausweb Poster Proceedings, Queensland, Australia}, volume~\bibinfo{volume}{15}, \bibinfo{organization}{Citeseer}, \bibinfo{year}{2001}, p.~\bibinfo{pages}{17}.
\bibitem[{Balog and De~Rijke(2007)}]{balog2007expertprofiling}
\bibinfo{author}{K.~Balog}, \bibinfo{author}{M.~De~Rijke},
\newblock \bibinfo{title}{Determining expert profiles (with an application to expert finding)},
\newblock in: \bibinfo{booktitle}{Proceedings of the 20th International Joint Conference on Artifical Intelligence}, IJCAI'07, \bibinfo{publisher}{Morgan Kaufmann Publishers Inc.}, \bibinfo{address}{San Francisco, CA, USA}, \bibinfo{year}{2007}, p. \bibinfo{pages}{2657–2662}.
\bibitem[{Balog et~al.(2012)Balog, Fang, de~Rijke, Serdyukov, and Si}]{balog2012expertise}
\bibinfo{author}{K.~Balog}, \bibinfo{author}{Y.~Fang}, \bibinfo{author}{M.~de~Rijke}, \bibinfo{author}{P.~Serdyukov}, \bibinfo{author}{L.~Si},
\newblock \bibinfo{title}{Expertise retrieval},
\newblock \bibinfo{journal}{Found. Trends Inf. Retr.} \bibinfo{volume}{6} (\bibinfo{year}{2012}) \bibinfo{pages}{127–256}. \URLprefix \url{https://doi.org/10.1561/1500000024}. \DOIprefix\doi{10.1561/1500000024}.
\bibitem[{Craswell et~al.(2005)Craswell, De~Vries, and Soboroff}]{craswell2005overview}
\bibinfo{author}{N.~Craswell}, \bibinfo{author}{A.~P. De~Vries}, \bibinfo{author}{I.~Soboroff},
\newblock \bibinfo{title}{Overview of the trec 2005 enterprise track.},
\newblock in: \bibinfo{booktitle}{Trec}, volume~\bibinfo{volume}{5}, \bibinfo{year}{2005}, pp. \bibinfo{pages}{1--7}.
\bibitem[{Soboroff et~al.(2008)Soboroff, Bailey, Craswell, and de}]{ian2008overview}
\bibinfo{author}{I.~Soboroff}, \bibinfo{author}{P.~Bailey}, \bibinfo{author}{N.~Craswell}, \bibinfo{author}{A.~de}, \bibinfo{title}{Overview of the trec 2007 enterprise track}, \bibinfo{year}{2008}. \URLprefix \url{https://tsapps.nist.gov/publication/get_pdf.cfm?pub_id=152187}.
\bibitem[{Deng et~al.(2008)Deng, King, and Lyu}]{deng2008formal}
\bibinfo{author}{H.~Deng}, \bibinfo{author}{I.~King}, \bibinfo{author}{M.~R. Lyu},
\newblock \bibinfo{title}{Formal models for expert finding on dblp bibliography data},
\newblock in: \bibinfo{booktitle}{2008 Eighth IEEE International Conference on Data Mining}, \bibinfo{year}{2008}, pp. \bibinfo{pages}{163--172}. \DOIprefix\doi{10.1109/ICDM.2008.29}.
\bibitem[{Balog et~al.(2007)Balog, Bogers, Azzopardi, de~Rijke, and van~den Bosch}]{balog2007broad}
\bibinfo{author}{K.~Balog}, \bibinfo{author}{T.~Bogers}, \bibinfo{author}{L.~Azzopardi}, \bibinfo{author}{M.~de~Rijke}, \bibinfo{author}{A.~van~den Bosch},
\newblock \bibinfo{title}{Broad expertise retrieval in sparse data environments},
\newblock in: \bibinfo{booktitle}{Proceedings of the 30th Annual International ACM SIGIR Conference on Research and Development in Information Retrieval}, SIGIR '07, \bibinfo{publisher}{Association for Computing Machinery}, \bibinfo{address}{New York, NY, USA}, \bibinfo{year}{2007}, p. \bibinfo{pages}{551–558}. \URLprefix \url{https://doi.org/10.1145/1277741.1277836}. \DOIprefix\doi{10.1145/1277741.1277836}.
\bibitem[{Berendsen et~al.(2013)Berendsen, De~Rijke, Balog, Bogers, and Van~den Bosch}]{berendsen2013assessment}
\bibinfo{author}{R.~Berendsen}, \bibinfo{author}{M.~De~Rijke}, \bibinfo{author}{K.~Balog}, \bibinfo{author}{T.~Bogers}, \bibinfo{author}{A.~Van~den Bosch},
\newblock \bibinfo{title}{On the assessment of expertise profiles},
\newblock \bibinfo{journal}{Journal of the American Society for Information Science and Technology} \bibinfo{volume}{64} (\bibinfo{year}{2013}) \bibinfo{pages}{2024--2044}.
\bibitem[{Mangaravite et~al.(2016)Mangaravite, Santos, Ribeiro, Gon\c{c}alves, and Laender}]{mangaravite2016lexr}
\bibinfo{author}{V.~Mangaravite}, \bibinfo{author}{R.~L. Santos}, \bibinfo{author}{I.~S. Ribeiro}, \bibinfo{author}{M.~A. Gon\c{c}alves}, \bibinfo{author}{A.~H. Laender},
\newblock \bibinfo{title}{The lexr collection for expertise retrieval in academia},
\newblock in: \bibinfo{booktitle}{Proceedings of the 39th International ACM SIGIR Conference on Research and Development in Information Retrieval}, SIGIR '16, \bibinfo{publisher}{Association for Computing Machinery}, \bibinfo{address}{New York, NY, USA}, \bibinfo{year}{2016}, p. \bibinfo{pages}{721–724}. \URLprefix \url{https://doi.org/10.1145/2911451.2914678}. \DOIprefix\doi{10.1145/2911451.2914678}.
\bibitem[{Borna et~al.(2024)Borna, Barry, Makarova, Parte, Haider, Sehgal, Leibovich, and Forte}]{sahar2024artificial}
\bibinfo{author}{S.~Borna}, \bibinfo{author}{B.~Barry}, \bibinfo{author}{S.~Makarova}, \bibinfo{author}{Y.~Parte}, \bibinfo{author}{C.~Haider}, \bibinfo{author}{A.~Sehgal}, \bibinfo{author}{B.~Leibovich}, \bibinfo{author}{A.~Forte},
\newblock \bibinfo{title}{Artificial intelligence algorithms for expert identification in medical domains: A scoping review},
\newblock \bibinfo{journal}{European Journal of Investigation in Health, Psychology and Education} \bibinfo{volume}{14} (\bibinfo{year}{2024}) \bibinfo{pages}{1182--1196}. \DOIprefix\doi{10.3390/ejihpe14050078}, \bibinfo{note}{publisher Copyright: {\textcopyright} 2024 by the authors.}
\bibitem[{Balog et~al.(2006)Balog, Azzopardi, and de~Rijke}]{balog2006formal}
\bibinfo{author}{K.~Balog}, \bibinfo{author}{L.~Azzopardi}, \bibinfo{author}{M.~de~Rijke},
\newblock \bibinfo{title}{Formal models for expert finding in enterprise corpora},
\newblock in: \bibinfo{booktitle}{Proceedings of the 29th Annual International ACM SIGIR Conference on Research and Development in Information Retrieval}, SIGIR '06, \bibinfo{publisher}{Association for Computing Machinery}, \bibinfo{address}{New York, NY, USA}, \bibinfo{year}{2006}, p. \bibinfo{pages}{43–50}. \URLprefix \url{https://doi.org/10.1145/1148170.1148181}. \DOIprefix\doi{10.1145/1148170.1148181}.
\bibitem[{Macdonald and Ounis(2006)}]{macdonald2006voting}
\bibinfo{author}{C.~Macdonald}, \bibinfo{author}{I.~Ounis},
\newblock \bibinfo{title}{Voting for candidates: adapting data fusion techniques for an expert search task},
\newblock in: \bibinfo{booktitle}{Proceedings of the 15th ACM International Conference on Information and Knowledge Management}, CIKM '06, \bibinfo{publisher}{Association for Computing Machinery}, \bibinfo{address}{New York, NY, USA}, \bibinfo{year}{2006}, p. \bibinfo{pages}{387–396}. \URLprefix \url{https://doi.org/10.1145/1183614.1183671}. \DOIprefix\doi{10.1145/1183614.1183671}.
\bibitem[{Fang and Zhai(2007)}]{fang2007prob}
\bibinfo{author}{H.~Fang}, \bibinfo{author}{C.~Zhai},
\newblock \bibinfo{title}{Probabilistic models for expert finding},
\newblock in: \bibinfo{editor}{G.~Amati}, \bibinfo{editor}{C.~Carpineto}, \bibinfo{editor}{G.~Romano} (Eds.), \bibinfo{booktitle}{Advances in Information Retrieval}, \bibinfo{publisher}{Springer Berlin Heidelberg}, \bibinfo{address}{Berlin, Heidelberg}, \bibinfo{year}{2007}, pp. \bibinfo{pages}{418--430}.
\bibitem[{Zhu et~al.(2006)Zhu, Song, R{\"u}ger, Eisenstadt, and Motta}]{zhu2006theopen}
\bibinfo{author}{J.~Zhu}, \bibinfo{author}{D.~Song}, \bibinfo{author}{S.~R{\"u}ger}, \bibinfo{author}{M.~Eisenstadt}, \bibinfo{author}{E.~Motta},
\newblock \bibinfo{title}{The open university at trec 2006 enterprise track expertsearch task},
\newblock in: \bibinfo{booktitle}{Fifteenth Text REtrieval Conference (TREC2006)}, \bibinfo{year}{2006}. \URLprefix \url{http://trec.nist.gov/pubs/trec15/papers/openu.ent.final.pdf}, \bibinfo{note}{proc. of The Fifteenth Text REtrieval Conference (TREC 2006), Gaithersburg, Maryland USA, National Institute of Standards and Technology, USA}.
\bibitem[{Van~Gysel et~al.(2016)Van~Gysel, de~Rijke, and Worring}]{gysel2016unsupervised}
\bibinfo{author}{C.~Van~Gysel}, \bibinfo{author}{M.~de~Rijke}, \bibinfo{author}{M.~Worring},
\newblock \bibinfo{title}{Unsupervised, efficient and semantic expertise retrieval},
\newblock in: \bibinfo{booktitle}{Proceedings of the 25th International Conference on World Wide Web}, WWW '16, \bibinfo{publisher}{International World Wide Web Conferences Steering Committee}, \bibinfo{address}{Republic and Canton of Geneva, CHE}, \bibinfo{year}{2016}, p. \bibinfo{pages}{1069–1079}. \URLprefix \url{https://doi.org/10.1145/2872427.2882974}. \DOIprefix\doi{10.1145/2872427.2882974}.
\bibitem[{Cifariello et~al.(2019)Cifariello, Ferragina, and Ponza}]{CIFARIELLO20191}
\bibinfo{author}{P.~Cifariello}, \bibinfo{author}{P.~Ferragina}, \bibinfo{author}{M.~Ponza},
\newblock \bibinfo{title}{Wiser: A semantic approach for expert finding in academia based on entity linking},
\newblock \bibinfo{journal}{Information Systems} \bibinfo{volume}{82} (\bibinfo{year}{2019}) \bibinfo{pages}{1--16}. \URLprefix \url{https://www.sciencedirect.com/science/article/pii/S0306437918302515}. \DOIprefix\doi{https://doi.org/10.1016/j.is.2018.12.003}.
\bibitem[{Robertson and Walker(1994)}]{robertson1994some}
\bibinfo{author}{S.~E. Robertson}, \bibinfo{author}{S.~Walker},
\newblock \bibinfo{title}{Some simple effective approximations to the 2-poisson model for probabilistic weighted retrieval},
\newblock in: \bibinfo{editor}{B.~W. Croft}, \bibinfo{editor}{C.~J. van Rijsbergen} (Eds.), \bibinfo{booktitle}{SIGIR '94}, \bibinfo{publisher}{Springer London}, \bibinfo{address}{London}, \bibinfo{year}{1994}, pp. \bibinfo{pages}{232--241}.
\bibitem[{Khattab and Zaharia(2020)}]{omar2020colbert}
\bibinfo{author}{O.~Khattab}, \bibinfo{author}{M.~Zaharia},
\newblock \bibinfo{title}{Colbert: Efficient and effective passage search via contextualized late interaction over bert},
\newblock in: \bibinfo{booktitle}{Proceedings of the 43rd International ACM SIGIR Conference on Research and Development in Information Retrieval}, SIGIR '20, \bibinfo{publisher}{Association for Computing Machinery}, \bibinfo{address}{New York, NY, USA}, \bibinfo{year}{2020}, p. \bibinfo{pages}{39–48}. \URLprefix \url{https://doi.org/10.1145/3397271.3401075}. \DOIprefix\doi{10.1145/3397271.3401075}.
\bibitem[{Louis et~al.(2024)Louis, Saxena, van Dijck, and Spanakis}]{louis2024modular}
\bibinfo{author}{A.~Louis}, \bibinfo{author}{V.~Saxena}, \bibinfo{author}{G.~van Dijck}, \bibinfo{author}{G.~Spanakis},
\newblock \bibinfo{title}{Colbert-xm: A modular multi-vector representation model for zero-shot multilingual information retrieval},
\newblock \bibinfo{journal}{CoRR} \bibinfo{volume}{abs/2402.15059} (\bibinfo{year}{2024}). \URLprefix \url{https://arxiv.org/abs/2402.15059}. \DOIprefix\doi{10.48550/arXiv.2402.15059}. \href{http://arxiv.org/abs/2402.15059}{{\tt arXiv:2402.15059}}.
\bibitem[{Wei et~al.(2022)Wei, Wang, Schuurmans, Bosma, Xia, Chi, Le, Zhou et~al.}]{wei2022chain}
\bibinfo{author}{J.~Wei}, \bibinfo{author}{X.~Wang}, \bibinfo{author}{D.~Schuurmans}, \bibinfo{author}{M.~Bosma}, \bibinfo{author}{F.~Xia}, \bibinfo{author}{E.~Chi}, \bibinfo{author}{Q.~V. Le}, \bibinfo{author}{D.~Zhou}, et~al.,
\newblock \bibinfo{title}{Chain-of-thought prompting elicits reasoning in large language models},
\newblock \bibinfo{journal}{Advances in neural information processing systems} \bibinfo{volume}{35} (\bibinfo{year}{2022}) \bibinfo{pages}{24824--24837}.
\bibitem[{Khattab et~al.(2023)Khattab, Singhvi, Maheshwari, Zhang, Santhanam, Vardhamanan, Haq, Sharma, Joshi, Moazam, Miller, Zaharia, and Potts}]{khattab2023dspy}
\bibinfo{author}{O.~Khattab}, \bibinfo{author}{A.~Singhvi}, \bibinfo{author}{P.~Maheshwari}, \bibinfo{author}{Z.~Zhang}, \bibinfo{author}{K.~Santhanam}, \bibinfo{author}{S.~Vardhamanan}, \bibinfo{author}{S.~Haq}, \bibinfo{author}{A.~Sharma}, \bibinfo{author}{T.~T. Joshi}, \bibinfo{author}{H.~Moazam}, \bibinfo{author}{H.~Miller}, \bibinfo{author}{M.~Zaharia}, \bibinfo{author}{C.~Potts},
\newblock \bibinfo{title}{Dspy: Compiling declarative language model calls into self-improving pipelines},
\newblock \bibinfo{journal}{arXiv preprint arXiv:2310.03714}  (\bibinfo{year}{2023}).
\bibitem[{Reimers and Gurevych(2019)}]{reimers-gurevych-2019-sentence}
\bibinfo{author}{N.~Reimers}, \bibinfo{author}{I.~Gurevych},
\newblock \bibinfo{title}{Sentence-{BERT}: Sentence embeddings using {S}iamese {BERT}-networks},
\newblock in: \bibinfo{editor}{K.~Inui}, \bibinfo{editor}{J.~Jiang}, \bibinfo{editor}{V.~Ng}, \bibinfo{editor}{X.~Wan} (Eds.), \bibinfo{booktitle}{Proceedings of the 2019 Conference on Empirical Methods in Natural Language Processing and the 9th International Joint Conference on Natural Language Processing (EMNLP-IJCNLP)}, \bibinfo{publisher}{Association for Computational Linguistics}, \bibinfo{address}{Hong Kong, China}, \bibinfo{year}{2019}, pp. \bibinfo{pages}{3982--3992}. \URLprefix \url{https://aclanthology.org/D19-1410}. \DOIprefix\doi{10.18653/v1/D19-1410}.
\bibitem[{Reimers and Gurevych(2020)}]{reimers-gurevych-2020-making}
\bibinfo{author}{N.~Reimers}, \bibinfo{author}{I.~Gurevych},
\newblock \bibinfo{title}{Making monolingual sentence embeddings multilingual using knowledge distillation},
\newblock in: \bibinfo{editor}{B.~Webber}, \bibinfo{editor}{T.~Cohn}, \bibinfo{editor}{Y.~He}, \bibinfo{editor}{Y.~Liu} (Eds.), \bibinfo{booktitle}{Proceedings of the 2020 Conference on Empirical Methods in Natural Language Processing (EMNLP)}, \bibinfo{publisher}{Association for Computational Linguistics}, \bibinfo{address}{Online}, \bibinfo{year}{2020}, pp. \bibinfo{pages}{4512--4525}. \URLprefix \url{https://aclanthology.org/2020.emnlp-main.365}. \DOIprefix\doi{10.18653/v1/2020.emnlp-main.365}.

\end{thebibliography}

\appendix


\section{Example annotation recommendations}\label{app:rec-examples}

Given the knowledge area \textsc{computer linguistics}, the top five item-to-item recommendations according to the PMI-based system are:
\small{
\begin{enumerate}[leftmargin=*]
    \item \textsc{talking computer}
    \item \textsc{automatic language analysis}
    \item \textsc{man-machine interaction}
    \item \textsc{algemene taalwetenschap (nl)}
    \item \textsc{speech technology}
\end{enumerate}
}

\noindent The top five recommendations according to the semantic embedding similarity method are:
\small{
\begin{enumerate}[leftmargin=*]
    \item \textsc{language and computers}
    \item \textsc{taalproductie door computers (nl)}
    \item \textsc{taaltechnologie en computers (nl)}
    \item \textsc{language technology and computers}
    \item \textsc{computer and grammar}
\end{enumerate}
}

Note that we use the Dutch topic name if no English name is available for the topic in Webwijs.
Now consider a self-selected expertise profile consisting of the following three topics:
\small{\begin{itemize}[leftmargin=*]
    \item \textsc{citizenship}
    \item \textsc{european law}
    \item \textsc{vreemdelingenrecht (nl)}
\end{itemize}}

\noindent The top 20 recommended annotations for this expert, as a result of the round-robin aggregation of both types of item-to-item recommenders, are:
\small{\begin{enumerate}[leftmargin=*]
    \item \textsc{voluntary work}
    \item \textsc{Nationality}
    \item \textsc{constitutional issues}
    \item \textsc{theory of European law}
    \item \textsc{Migration Law}
    \item \textsc{competition law}
    \item \textsc{foundations of developments in european law}
    \item \textsc{Refugee law}
    \item \textsc{vrijwilligersorganisaties (nl)}
    \item \textsc{inburgering van allochtonen (nl)}
    \item \textsc{legal principles}
    \item \textsc{European Administrative Law}
    \item \textsc{administrative law}
    \item \textsc{international law}
    \item \textsc{urban governance}
    \item \textsc{vreemdelingenbeleid (NL)}
    \item \textsc{rechtsbescherming (NL)}
    \item \textsc{european social law}
    \item \textsc{municipal law}
    \item \textsc{maatschappelijke organisaties (NL)}
\end{enumerate}}

\end{document}